\def\be{\begin{equation}}
\def\ee{\end{equation}}
\def\kpc{\;{\rm kpc} }
\def\pc{\;{\rm pc} }
\def\pcm3{\;{\rm pc}^{-3} }

\def\msun{{\rm M}_{\rm \odot}}

\def\kms{\;{\rm km} {\rm s}^{-1}  }
\def\sigmalos{\sigma_{\rm los}}
\def\sigmaz{{\sigma_z}}
\def\sec{{\rm sec}}

\def\hd{h_{\rm d}}
\def\hs{h_{\rm s}}
\def\hv{h_{\rm v}}

\def\fshroud{f_{\rm s}}
\def\fv{f_{\rm v}}

\def\that{{\hat t}}
\def\gshroud{f_{\rm s}}
\def\fhalo{f_{\rm h}}
\def\mdark{m_{\rm dark}}

\def\phihalo{\phi_{\rm h}}
\def\Lshroud{L_{\rm s}}
\def\Lhalo{L_{\rm h}}

\def\spose#1{\hbox to 0pt{#1\hss}}
\def\lta{\mathrel{\spose{\lower 3pt\hbox{$\sim$}}
    \raise 2.0pt\hbox{$<$}}}
\def\gta{\mathrel{\spose{\lower 3pt\hbox{$\sim$}}
    \raise 2.0pt\hbox{$>$}}}

\documentstyle[12pt,aasms4,epsf,rotate]{article}

\begin{document}

\title{Is the Large Magellanic Cloud a Large Microlensing Cloud?}

\author{N. Wyn Evans and Eamonn Kerins}
\affil{Theoretical Physics, Department of Physics, 1 Keble Rd, Oxford,
OX1 3NP, UK}

\begin{abstract} 
An expression is provided for the self-lensing optical depth of the
thin LMC disk surrounded by a shroud of stars at larger scale
heights. The formula is written in terms of the vertical velocity
dispersion of the thin disk population. If tidal forcing causes $\sim
1-5 \%$ of the disk mass to have a height larger than 6 kpc and $\sim
10-15 \%$ to have a height above 3 kpc, then the self-lensing optical
depth of the LMC is $\sim 0.7 - 1.9 \times 10^{-7}$, which is within
the observational uncertainties. The shroud may be composed of bright
stars provided they are not in stellar hydrodynamical equilibrium.
Alternatively, the shroud may be built from low mass stars or compact
objects, though then the self-lensing optical depths are overestimates
of the true optical depth by a factor of $\sim 3$.

The distributions of timescales of the events and their spatial
variation across the face of the LMC disk offer possibilities of
identifying the dominant lens population. We use Monte Carlo
simulations to show that, in propitious circumstances, an experiment
lifetime of $\lta 5$ years is sufficient to decide between the
competing claims of Milky Way halos and LMC lenses. However, LMC disks
can sometimes mimic the microlensing properties of Galactic halos for
many years and then decades of survey work are needed for
discrimination. In this case observations of parallax or binary caustic
events offer the best hope for current experiments to deduce the
dominant lens population.  The difficult models to distinguish are
Milky Way halos in which the lens fraction is low ($\lta 10 \%$)
and fattened LMC disks composed of lenses with a typical mass of low
luminosity stars or greater. A next-generation wide-area microlensing
survey, such as the proposed ``SuperMACHO'' experiment, will be able
to distinguish even these difficult models with just a year or two of
data.
\end{abstract}


\section{INTRODUCTION}

The location of the microlensing events towards the Large Magellanic
Cloud (LMC) is a matter of controversy. Alcock et al. (1997a) assert
that the lensing population lies in the Galactic halo and comprises
perhaps $\sim 50 \%$ of its total mass. Early suggestions that the LMC
may provide the bulk of the lenses were made by Sahu (1994) and Wu
(1994), and this location is favored by the data on the binary caustic
crossing events (Kerins \& Evans 1999). One of the main obstacles to
general acceptance of this idea has been the sheer number of observed
lensing events, which appear to be too great to be accommodated by the
LMC alone. The experimental estimate of the microlensing optical depth
$\tau$ towards the LMC is $2.1^{+1.3}_{-0.8} \times 10^{-7}$ (e.g.,
Bennett 1998).  This is substantially greater than the optical depth
of simple tilted disk models of the LMC. For example, Gould's (1995)
ingenious calculation involving the virial theorem sets the
self-lensing optical depth of the LMC disk as $\sim 1 \times 10^{-8}$.
Section 2 of this paper generalizes Gould's analysis to provide upper
limits on the self-lensing optical depth of thick models of the LMC
disk. These values are smaller than, but of comparable magnitude to,
the observations.  So, it is reasonable to suggest that the
microlensing signal may come either from a fattened LMC disk or a
Milky Way halo only partly composed of lensing objects.  The timescale
distributions and the geometric pattern of events across the face of
the LMC disk will of course be different in these two cases.  The
timescales of events for the same mass functions will be longer for
lenses in the LMC as compared to those in the Milky Way halo as the
lower velocity dispersion of the LMC outweighs the effects of the
smaller Einstein radii. However, the use of the timescales as a
discriminant is spoiled by the fact that there is no reason why the
Milky Way halo and the LMC should have the same mass function.  A more
hopeful indicator may be the distribution of events across the face of
the LMC disk. If the lenses lie in the Milky Way halo, the events will
trace the surface density of the LMC, whereas if the lenses lie in the
Clouds, the events will be more concentrated towards the dense bar and
central regions, scaling like the surface density squared.  How long
will it take to distinguish between the two possibilities?  Section 3
develops a maximum likelihood estimator that incorporates all the
timescale and positional information to provide the answer to this
question, both for the existing surveys like MACHO and for the next
generation experiments like SuperMACHO (Stubbs 1998).  Finally,
Section 4 evaluates the strategies by which the riddle of the location
of the lenses may be solved.

\section{OBESE MAGELLANIC DISKS}

Gould's (1995) limit relates the self-lensing optical depth of any thin
disk to its vertical velocity dispersion $\sigmaz$ via
\begin{equation}
\tau = 2 {\sigmaz^2\over c^2} \sec^2 i
\end{equation}
where $c$ is the velocity of light and $i$ is the inclination angle.
Taking the observed velocity dispersion of CH stars as $\sim 20 \kms$
(Cowley \& Hartwick 1991) and the inclination angle $i = 27^\circ$ (de
Vaucouleurs \& Freeman 1973), Gould argued that the self-lensing
optical depth of the LMC disk is likely to be $\sim 1 \times 10^{-8}$,
which is some 20 times smaller than the observations.  As Gould's
derivation depends only on the Poisson and Jeans equations for highly
flattened geometries, the formula is clearly irreproachable. How could
it be yielding misleading results as to the self-lensing optical
depth?  Consider a thought experiment in which a very thin disk is
gradually surrounded by a flattened shell of matter, bounded by two
similar concentric ellipsoids (a homoeoid). By Newton's theorem, the
attraction at any internal point of a homoeoid vanishes. So, the
introduction of the homoeoid leaves the velocity dispersion in the
thin disk quite unchanged.  But, the self-lensing optical depth is
strongly enhanced.  In applying Gould's formula, we must be {\it very}
careful not to use the velocity dispersion of the thin disk, but
rather the mass-weighted velocity dispersion of the entire
configuration -- otherwise we will obtain a misleadingly small answer.
It is clearly worthwhile extending the calculations to more
sophisticated, structure-rich models of the LMC, in which the
self-lensing optical depth is written in terms of the velocity
dispersion of the thin disk population (which is directly accessible
to observations) instead of the mass-weighted velocity dispersion.

Let us now derive formulae for the self-lensing optical depth of an
ensemble of $n$ exponential disks, each with a different scale height
$h_i$, mid-plane density $\rho_i$ and column density $\Sigma_i = 2
\rho_i h_i$. Clearly, this is a very idealized representation of the
LMC, although similar models of the Milky Way disk have already proved
useful (c.f., Gould 1989). The vertical density law for the disk is
\begin{equation}
\rho(z) =  \sum_{i=1}^n \rho_i \exp \left( -{|z|\over h_i} \right).
\end{equation}
The relationship between height $z$ and potential $\phi$ is given by
solving Poisson's equation in the form appropriate for a flattened
geometry (see Binney \& Tremaine 1987, chap 2).  Gould (1995) shows
that the self-lensing optical depth of any thin disk with total column
density $\Sigma$ is
\begin{equation}
\label{gouldseq}
\tau = {2 \pi G \Sigma \sec^2 i \over c^2} \int_0^\infty dz\{ 1-
[G(z)]^2\},\qquad\qquad G(z) = {2\over \Sigma}\int_{0}^z dy \rho(y)
\end{equation}
For our ensemble of exponential disks we have
\begin{equation}
G(z) = 1 - \sum_{i=1}^n F_i \exp \left( -{|z| \over h_i} \right). 
\end{equation}
where $F_i$ is the mass fraction in each population. The self-lensing
optical depth is now entirely analytic and given by
\begin{equation}
\tau = 2 {\sigma_1^2\over c^2} \sec^2 i \times { 1\over F_1 h_1}
\left[ {4\over 3} \sum_{i=1}^n F_i h_i - {2\over 3}
\sum_{i,j=1}^n {F_i F_j h_i h_j \over h_i + h_j} \right],
\label{eq:multiexp}
\end{equation}
The formula has been written in terms of $\sigma_1$, which is the
vertical velocity dispersion of the thinnest disk population only. It
is the line-of-sight velocity dispersion of the youngest, thinnest
populations in the LMC that are observationally reasonably
well-determined. The Jeans equation has been solved under the
assumption that the thinnest disk dominates the gravitational
potential near the mid-plane. Our formula only assumes that the
thinnest population is in equilibrium. It makes no assumptions as to
the relationship between velocity dispersion and height for the
thicker populations.  It is therefore the appropriate formula for an
equilibrium thin disk surrounded by dispersed and patchy populations
of stars. Let us note that (\ref{eq:multiexp}) really estimates the
value of the optical depth near the LMC center (as the radial
structure of the disk is ignored).  The assumption that the disks are
exponential rather than completely isothermal (that is, sech-squared)
causes our estimates to be on the low side.  The assumption that the
line-of-sight velocity dispersion $\sigmalos$ is roughly equal to the
vertical velocity dispersion $\sigma_z$ causes our estimates to be on
the high side. This correction factor depends on the uncertain shape
of the velocity dispersion tensor in the LMC thin disk. If the
velocity dispersions in the disk are well-approximated by epicyclic
theory, then $\sigma_\phi^2 \sim \sigma_z^2 \sim 0.45 \sigma_R^2$
(Binney \& Tremaine, 1987, p. 199). In this case, $\sigmalos^2$
overestimates $\sigma_z^2$ by $\sim ( 1+ 0.6 \sin^2 i)$.  Finally, in
the limit of a single, thin disk ($n=1$), our
result~(\ref{eq:multiexp}) reduces to Gould's original formula, as it
should.

Weinberg (1999) has described self-consistent simulations of the tidal
forcing of the Magellanic Clouds by the Milky Way galaxy. He shows
that the effect of this tidal heating is to fatten the structure of
the LMC. He reports that $\sim 1 \%$ of the disk mass has a height
larger than 6 kpc (which we will call ``the veil'') and $\sim 10 \%$
above 3 kpc (``the shroud'').  Let us devise a three component model
of the LMC, composed of a massive thin disk surrounded by an
intermediate shroud and an extended veil.  To model the LMC, let us
take the scale height of the thin disk as $\hd \sim 300
\pc$ (Bessell, Freeman \& Wood 1986). The vertical velocity dispersion
of the stars in this disk is $\sim 30 \kms$.  The scale heights of the
shroud $\hs$ and the veil $\hv$ are $3 \kpc$ and $6 \kpc$
respectively. As suggested by Weinberg's (1999) calculation, we put
$10 \%$ of the mass in the shroud and $1 \%$ in the veil. De
Vaucouleurs \& Freeman (1973) estimated the inclination angle of the
LMC to be $27^\circ$ by assuming that the optical and 21 cm HI
isophotes should be circular.  This is not likely to be a good
approximation for such an irregular structure like the LMC, and so
this widely-used value of the inclination angle is at least open to
some doubt.  More recently, evidence from detailed fitting of the
surface photometry (excluding the star forming regions) and from the
low frequency radio observations (which are less sensitive to local
effects) suggest a higher value of the inclination angle of the main
disk of $i \sim 45^\circ$ (see e.g., Alvarez, Aparici \& May 1987;
Bothun \& Thomson 1988). Westerlund (1997) reviews all the evidence
and argues that this higher value of the inclination is most
likely. We will consider both possibilities.  When $10 \%$ of the mass
is in the shroud and $1 \%$ in the veil, the self-lensing optical
depth is $0.7 \times 10^{-7}$ if $i = 27^\circ$ and $1.1 \times
10^{-7}$ if $i = 45^\circ$.  Figure 1 shows how the self-lensing
optical depth varies as the mass fractions in the shroud and the veil
are changed. Marked on Figure 1 are the contours corresponding to the
best observational estimate of $2.1 \times 10^{-7}$, together with the
$1 \sigma$ and $2\sigma$ lower limits.  If the mass fractions are
increased to $15 \%$ and $5\%$ respectively, then the optical depth is
$1.2 \times 10^{-7}$ if $i = 27^\circ$ and $1.9 \times 10^{-7}$ if $i
= 45^\circ$.  These values are comparable to the observed optical
depth of $2.1^{+1.3}_{-0.8} \times 10^{-7}$ (Bennett 1998).  On moving
to the larger inclination, the assumption that the line-of-sight
dispersion is roughly equal to the vertical velocity dispersion
becomes less valid. Using our earlier correction based on epicyclic
theory, some $\sim 15 \%$ of the increase in the optical depth on
moving to the larger inclination of $45^\circ$ is spurious.  However,
the important conclusion to draw from these calculations is that it
requires comparatively little luminous material at higher scale
heights above the LMC thin disk to give a substantial boost to the
optical depth.

There is one obvious difficulty with this suggestion.  There are no
visible tracers in the LMC with a velocity dispersion greater than $33
\kms$ (Hughes, Wood \& Reid 1991; Westerlund 1997).  If in
equilibrium, any luminous material belonging to disks with scale
heights of $3\kpc$ or $6\kpc$ must have a larger velocity dispersion
than observed.  For example, the tidal heating mechanism advocated by
Weinberg (1999) must produce some visible hot tracers.  The stars that
are heated are expected to have the same luminosity function as those
that remain in the thin disk. There are two possible loopholes in this
line of argument. First, it might be possible for a metal-rich, old
population with a large velocity dispersion to have eluded detection.
Second, the relationship between scale height and velocity dispersion
applies only to steady-state equilibrium models. If this is not the
case, then it may be possible for populations to be dispersed at
larger heights above the LMC thin disk than suggested by their
vertical velocity dispersion. It is worth cautioning that equilibrium
models of the LMC may be a poor guide to interpreting the
kinematics. In particular, no equilibrium models of galaxies with
off-centered bars are presently known, either analytically or as the
endpoints of N body experiments. If both these loopholes are closed,
then the last possibility is that any lenses in the larger scale
height populations must be dark or at very least dim -- perhaps low
mass stars or compact objects. This is difficult to rule out, although
there are no obvious natural mechanisms to produce such components. In
this case, our self-lensing formula~(\ref{eq:multiexp}) will
overestimate the microlensing optical depth, as the population of
lenses and sources do not coincide. It should be replaced by
\begin{equation}
\tau =  {2 \over 3} {\sigma_1^2\over c^2} \sec^2 i \times { 1\over F_1 h_1}
\sum_{i=1}^n F_i h_i 
\label{eq:nonselflens}
\end{equation}
For the same mass fractions $F_i$, the optical
depths~(\ref{eq:nonselflens}) are reduced by a typical factor of $\sim
3$ from our earlier self-lensing estimates~(\ref{eq:multiexp}).
Aubourg et al. (1999) and Salati et al. (1999) have recently advanced
models of the LMC surrounded by swathes of low mass stars and
suggested that they could provide most of the observed microlensing
optical depth, although others have contested this (e.g., Gyuk, Dalal
\& Griest 1999).

\section{THE LOCATION OF THE LENSES}

Can the positions and timescales of the microlensing events be used to
determine whether the dominant lens population lies in the LMC or in
the Milky Way halo?  The Bayesian likelihood estimator employed by
Alcock et al. (1997a) can be extended to consider lenses from multiple
galactic components distributed over a finite solid angle. For an
experiment of lifetime $T$ in which $N(T)$ events are observed with
Einstein diameter crossing durations $\that_i$ and Galactic coordinates
$l_i,b_i$ ($i = 1 \ldots N$), one can ascribe a likelihood $L$, where
\be
      \ln L(f_{1 \ldots n},\phi_{1 \ldots n}) = -\sum_{j=1}^n f_j
      {\cal N}(\phi_j,T) + \sum_{i = 1}^{N(T)} \ln \left[ \sigma(l_i,b_i)
      {\cal E}(\that_i,l_i,b_i,T) \sum_{j=1}^n f_j \frac{ {\rm d}
      \Gamma(\phi_j,l_i,b_i)}{{\rm d}\that_i} \right], \label{like}
\ee
to a galactic model comprising $j = 1 \ldots n$ components, each
component being characterised by a lens fraction $f_j$ and mass
function $\phi_j$. In the above formula, $\Gamma$ is the theoretical
event rate, ${\cal E}$ is the detection efficiency, $\sigma$ is the
number of sources per unit solid angle and
\be
      {\cal N}(\phi_j,T) = T \int \int \int \sigma(l,b) {\cal
      E}(\that,l,b,T)\frac{{\rm d} \Gamma(\phi_j,l,b)}{{\rm d}\that} \,
      {\rm d}\that\, {\rm d}l \, {\rm d}(\sin b) \label{num}
\ee
is the number of events predicted for component $j$ when $f_j =
1$. The spatial variation of microlensing events has been studied
before by Gyuk (1999), though using the optical depth and rate
rather than the timescales (and with the emphasis on the
inner Galaxy). 

Let us set up two competing models. In the first, the Milky Way halo
provides the dominant lens population, although there is some residual
contribution from the stars in the LMC disk and bar. In the second,
there is no Milky Way halo and the LMC disk and bar are augmented by the
existence of an enveloping shroud and veil, so that all the lenses
reside close to or in the LMC. The density laws describing the
components are summarised in Table~\ref{table:wyn}.  In both cases,
the LMC disk and bar are populated with lens masses $m$ drawn from the
ordinary stellar disk population. The broken power-law
\begin{eqnarray}
      \phi_{\rm LMC} & \propto & m^{\gamma} \quad \quad 
      (m_{\rm{L}} = 0.08~\msun \leq m \leq m_{\rm U} = 10~\msun),
      \nonumber \\
      \gamma & = & \left\{ \begin{array}{ll}
       -0.75 & (m_{\rm L} \leq m < 0.5~\msun) \\
       -2.2 & (0.5~\msun \leq m \leq m_{\rm U})
       \end{array} \right.
\end{eqnarray}
describes the LMC stellar mass function (c.f., Hill, Madore \&
Freedman 1994; Gould, Bahcall \& Flynn 1997). For our Milky Way halo,
we adopt a $\delta$-function
   \be
      \phihalo \propto \frac{1}{m} \delta( m - \mdark )
   \ee
as characterising the lens mass. For the competing LMC-only model,
there is an extended shroud and veil (hereafter collectively referred
to simply as the shroud) enveloping the LMC stellar disk and bar. For
simplicity, let us investigate the case in which the shroud consists
primarily of dark lenses (either remnants or low-mass stars).  Since
the Milky Way halo and LMC shroud populations are both dark, we always
make comparisons assuming the same lens mass $\mdark$.  For this
calculation, we make the simplifying assumption that the LMC is
virialized, so that any increase in the mass of the shroud implies a
corresponding increase in its velocity dispersion. This is important
because changes in the velocity dispersion affect the derived lens
timescale distribution.  Suppose the ratio of the disk to shroud
masses is originally $r$.  Then if the mass of our shroud is increased
by a factor $\gshroud$, the virial theorem indicates that the velocity
dispersion increases by a factor $f_{\sigma} =
\sqrt{(\gshroud+ r)/(1+r)}$. We must also make the corresponding
transformations $\that \rightarrow f_{\sigma}^{-1} \that$ and ${\rm
d}\Gamma/{\rm d}\that \rightarrow f_{\sigma}^2 \gshroud {\rm
d}\Gamma/{\rm d}\that$.

Let us proceed by simulating microlensing experiments over a range of
lifetimes $T$. We assume the Milky Way halo is an isothermal spherical
halo of amplitude $v_0 = 220 \kms $. A fraction $\fhalo$ of the halo
comprises lenses of mass $\mdark$. This provides us with our input
model with which to generate ``observed'' events. The expected number
of events for an experiment of lifetime $T$ is simply ${\cal N}(T) =
{\cal N}(\phi_{\rm LMC},T) + \fhalo {\cal N}(\phihalo,T)$, where
${\cal N}(\phi_{\rm LMC},T)$ and ${\cal N}(\phihalo,T)$ are obtained
from eqn~(\ref{num}). We then generate a Poisson realisation $N(T)$
for the number of observed events. We approximate the current
generation of microlensing surveys by an ideal experiment which
monitors the central $3^{\circ} \times 3^{\circ}$ of the LMC. For each
event a location is generated from within this region using the
distribution
   \be P(l,b|T) \propto \sigma(l,b) \int {\cal E}(\that,l,b,T) \left[
      \frac{{\rm d} \Gamma(\phi_{\rm LMC},l,b)}{{\rm d}\that} + \fhalo
      \frac{{\rm d}\Gamma(\phihalo,l,b)}{{\rm d}\that} \right] \, {\rm
      d}\that, \label{pos} \ee
which traces the event number density as a function of position. The
Einstein diameter crossing time $\that$ is generated from the
distribution
   \be
      P(\that|l,b,T) \propto {\cal E}(\that,l,b,T) \left[ \frac{{\rm
      d}\Gamma(\phi_{\rm LMC},l,b)}{{\rm d}\that} + \fhalo
      \frac{{\rm d}\Gamma(\phihalo,l,b)}{{\rm d}\that} \right].
   \ee
The detection efficiency ${\cal E}$ is not just a function of $\that$ and
$T$, but also Galactic coordinates $l$ and $b$. The spatial dependency
of ${\cal E}$ has not yet been assessed by any of the current
experiments and is inevitably experiment-specific. In the following
analysis we consider an idealized microlensing survey in which the
spatial dependency is sufficiently weak to be neglected. This is not a
good assumption for the current LMC microlensing surveys which do not
observe all regions with the same frequency, but the method we present
is general and can be used to take account of spatial variations in
efficiencies when these become available.  As microlensing experiments
continue, they become more sensitive to longer duration
events. However, the efficiency ${\cal E}$ does not approach unity
because of photometric limits imposed by the observing
conditions. Instead one might anticipate, say, a limiting efficiency
${\cal E}_{\rm max} \approx 0.5$. We propose the following model for
the time evolution of the efficiency for our ideal experiment:
   \be
      {\cal E} =  \left\{ \begin{array}{ll}
      \max[0,{\cal E}_{\rm short}(\that)] & (\that < \that_{\rm peak}) \\
      \max[0,{\cal E}_{\rm short}(\that_{\rm peak})] \exp \{ -[\log(\that_{\rm
      peak}/\that)/0.5]^2 \} & (\that \geq \that_{\rm peak})
      \end{array}\right. \label{eq:eff1}
   \ee
where
\be 
      \that_{\rm peak} = 0.12 \, T \qquad {\cal E}_{\rm short} = 
      \min \{ {\cal E}_{\rm max} , 0.2 [\log (\that/\mbox{days})
      -0.38] \} \qquad {\cal E}_{\rm max} = 0.5. \label{eq:eff2}
\ee
Here, $\that_{\rm peak}$ is the Einstein diameter crossing time at
which the efficiency peaks, which of course depends on the experiment
lifetime $T$.  As Figure~\ref{fig:eff} shows, the model (dashed lines)
provides an excellent approximation of the Alcock et al. (1997a)
2.1-year efficiencies (solid line). It is also broadly consistent with
provisional 4-year MACHO efficiency estimates (Sutherland 1999).
Note from Figure~\ref{fig:eff} that the limiting efficiency ${\cal
E}_{\rm max}$ is not reached until $T \simeq 20$ years, much longer
than the nominal lifetime of the MACHO experiment. Let us emphasize
that this model is only a plausible representation of how the
efficiencies for the current generation of microlensing experiments
might evolve.

We can now use simulated datasets to compute likelihoods for any
desired theoretical model via eqn~(\ref{like}).  For the dataset, we
calculate the likelihood $\Lhalo$ for the input (true) halo, LMC disk
and bar parameters. Let the likelihood of the competing model of a
shrouded LMC disk and bar be $\Lshroud$.  The ratio $\Lshroud /\Lhalo$
then provides a direct measure of the preference of the dataset for
the (true) halo model or (false) shroud model. Given just these two
alternatives, we can define a discrimination measure
        \be D =
        \frac{\Lhalo}{\Lshroud+\Lhalo}
        \ee
which is the probability, given the data, that the halo rather than
the shroud, represents the underlying model.  Individual datasets can
be misleading, so we generate a large ensemble of datasets for every
experiment lifetime $T$.  (Specifically, we use either $10^5$
datasets or a cumulative total of $3 \times 10^6$ events, whichever is
reached first).  From the resulting distribution of $D$ values, it is
possible to assess not just the degree of discrimination for a
particular dataset between the input and comparison model, but also
the likelihood of obtaining a dataset with at least that level of
discrimination.

In Figure~\ref{fig:res}, we plot $D_{95}$ (that is, the $95\%$ lower
limit on $D$) computed from the ensemble of simulated datasets, for a
variety of input and comparison models (all assuming $\gshroud = 1$)
for experiments with a lifespan of up to 20 years. The figure clearly
illustrates how much longer it takes to distinguish between the
competing models for smaller halo fractions and larger lens masses.
For halo fractions $\fhalo \ga 0.3$, we expect our experiment to
clearly distinguish between the two models after about 5~years if
$\mdark \la 0.5~\msun$. The amount of time $T$ required to decisively
reject the shroud model is about twice as long for lens masses $\mdark
= 0.5~\msun$ than for $\mdark = 0.1~\msun$, and this is due to the
larger number of events typically observed for the lower mass
lenses. If our ideal experiment is indicative of the progress of the
MACHO survey, it seems that even after 10~years the experiment may
still be unable to clearly distinguish between the halo and shrouded
LMC models if $\fhalo \sim 0.1$ and $\mdark \ga 0.1~\msun$.

Table~\ref{table:eamonn} shows the experimental lifetime $T$, in
years, required to constrain $D_{95} > 0.95$, at which point $95\%$ of
datasets clearly reject the shroud model. The limits displayed in
Figure~\ref{fig:res} are summarised in columns~2--4 of
table~\ref{table:eamonn}. Columns~5--7 show the equivalent limits if
one employs a likelihood statistic that does not take into account the
spatial distribution of the events. Columns~8--10 are for a shroud
mass factor half as large as assumed in figure~\ref{fig:res}.  For
columns~5--7, we have assumed that the timescale distribution at all
locations is the same as the distribution for the line of sight
through the LMC centre.  We see that the spatial distribution of
events becomes an increasingly important discriminant for halos of
lower lens fraction and lenses of larger mass. In the case where
$\fhalo = 0.1$, the incorporation of the angular distribution of
events into the likelihood statistic greatly enhances the sensitivity
of the analysis to the lens population. In this case the number of
generated events are similar for the competing halo and shroud models,
and so the spatial distribution becomes an important discriminatory
factor.  Comparing columns~2--4 with columns~8--10, where a shroud
mass factor $\gshroud = 0.5$ is assumed, we see that less massive LMC
disks are easier to distinguish. The overall constraints for $\gshroud
= 0.5$ are stronger for a given $T$ than those for $\gshroud = 1$
because the halo events always outnumber LMC events. The relative
constraints for different $\fhalo$ and $\mdark$ are similar for both
values of $\gshroud$; LMC and halo models with $\mdark =
0.5~\msun$ are about twice as difficult to differentiate as those
with $\mdark = 0.1~\msun$.

Stubbs (1998) has proposed a next generation microlensing survey
(provisionally dubbed ``SuperMACHO'') capable of detecting events at a
rate at least an order of magnitude greater than current
experiments. Gould (1999) finds that coverage of the whole LMC disk is
the key to maximizing the returns of such a survey. In
Figure~\ref{fig:sm} we compare the discrimination capability of
SuperMACHO with that of current surveys, assuming that SuperMACHO
commences nine years after the current surveys, and that the current
experiments are continued through the next decade (in reality, the
current surveys are scheduled to terminate in the next few
years). Let us assume that the SuperMACHO angular coverage will be as
suggested by Gould (1999), namely $11^{\circ} \times 11^{\circ}$
centered on the LMC bar, that the number of detected events will be ten
times greater than current yields, and that the detection efficiency
evolves according to equations~(\ref{eq:eff1}) and (\ref{eq:eff2}). In
reality the SuperMACHO detection efficiency is likely to be
qualitatively different than for the current experiments because many
of the central fields will be strongly blended.

In Figure~\ref{fig:sm} we have re-plotted the $95\%$ limit on the
discriminatory power ($D_{95}$) of current surveys for the case
$\fhalo = 0.1$, $\mdark = 0.5~\msun$ (solid line). This time we plot
$D_{95}$ against epoch rather than experiment lifetime. We adopt
1992.5 as the start of the current surveys (it actually corresponds to
the start of the MACHO survey) with the SuperMACHO survey, shown by
the dashed line, is assumed to start in 2001. Whilst current surveys
would take 20 years to distinguish clearly between LMC and halo
populations for this model, SuperMACHO takes only 18 months to reach
the same level of discrimination. SuperMACHO will surpass the
sensitivity of current surveys within a year of starting (if it indeed
starts on the assumed date). A survey along the lines of SuperMACHO
represents one of the best ways to discriminate statistically between
halo and LMC lens populations in the next few years.

\section{CONCLUSIONS}

It is possible to build models of the Large Magellanic Cloud (LMC)
with microlensing optical depths that are comparable to, although
lower than, the observations. Such models are fatter than is
conventional, with material extending to scale heights of $\sim 6$ kpc
above the plane of the LMC disk, as is suggested by Weinberg's (1999)
numerical simulations of the evolution of the LMC in the tidal field
of the Milky Way. This paper has derived the formula for the
self-lensing optical depth of an equilibrium thin disk surrounded by
stars dispersed at greater scale heights. As a shorthand, we call such
material the shroud, even though its distribution may be quite
patchy. When $\sim 10 \%$ of the total column density is in the
shroud, the self-lensing optical depth is typically between $0.7
\times 10^{-7}$ and $1.1 \times 10^{-7}$.  The self-lensing optical 
depth rises to between $1.2 \times 10^{-7}$ and $1.9 \times 10^{-7}$
when $\sim 20 \%$ of the column density is in the shroud.  These
figures should be compared to the observational estimate of
$2.1^{+1.3}_{-0.8} \times 10^{-7}$.  Provisional estimates using the
4-year dataset suggest that the optical depth may be lower (Sutherland
1999). Additionally, the difficulty of reproducing the high optical
depths reported by both Udalski et al. (1994) and Alcock et
al. (1997b) towards the Galactic Center using barred models of the
inner Galaxy (e.g., H\"afner et al. 1999) hints at a possible
systematic over-estimate afflicting the experimental values.  Clearly,
the suggestion that almost all the microlensing events emanate at or
close to the LMC cannot be dismissed lightly.

The difficulty with fattening the LMC disk is that there are no known
LMC populations with a line of sight velocity dispersion exceeding $33
\kms$ (Hughes et al. 1991). Stars in equilibrium in a thick disk 
with a scale height of $3 \kpc$ typically possess a larger velocity
dispersion than this. One possibility is that the shroud stars belong
to an old, metal-rich population that could have evaded
detection. More likely, perhaps, is that the material in the shroud is
not in a steady-state at all. Its spatial distribution may be quite
patchy, making it difficult to pick out against the bright central
bar. A final option is that the shroud is composed of dark or dim
material, such as low mass stars or compact objects (c.f. Aubourg et
al. 1999). Self-lensing optical depths then overestimate the true
optical depth by a factor of $\sim 3$, though this may be partly
compensated by increasing the mass fraction in the shroud. The idea is
tantamount to enveloping the LMC in its own dark halo. So, a shrouded
LMC may not dispense with the need for compact dark matter. It merely
re-locates it from the Milky Way halo to the LMC, though of course a
much lower total mass budget in compact objects is implied. A dark
shroud is difficult to rule out, although there is no obvious way to
arrange the low mass stars or compact objects around the LMC thin
disk.

It is natural to hope that the spatial distribution of events across
the face of the LMC disk and the timescale information can be used to
identify the main location of the lenses. In some circumstances, an
experiment lifetime of $\lta 5$ years is sufficient to decide between
the competing claims of Milky Way halo lenses and LMC lenses. However,
there is an awkward r\'egime in which fattened LMC disks can mimic
anorexic halos and several decades of survey work are needed for
discrimination.  The difficult models to distinguish are Milky Way
halos in which the lens fraction is very low ($\fhalo \lta 0.1$) and
obese LMC disks composed of lenses with a typical mass of low
luminosity stars or greater, $\mdark \ga 0.1~\msun$. This suggests
that the timescales and the geometric distribution of the microlensing
events may not be sufficient for an unambiguous resolution of the
puzzle of the origin of the lenses within the lifetime of the current
surveys.

One suggested approach to this problem is to employ a much more
sensitive microlensing survey covering the whole LMC disk, not just
the regions around the bar. The proposed ``SuperMACHO'' survey (Stubbs
1998) should be able to discriminate between even anorexic halos and
fattened LMC disks within 18 months of starting. So, the commencement
of a program like SuperMACHO represents one of the most promising ways
to answer this question in the next few years.  In the meantime, we
may still hope to differentiate between the lens locations using data
from binary caustic crossing events and from the presence or absence
of parallax events. As Kerins \& Evans (1999) have already argued, the
former are a particularly powerful diagnostic. If the next binary
caustic crossing event has a high projected velocity, then this
securely establishes a lensing component in the Milky Way halo. If the
next binary caustic crossing event has a low projected velocity, then
-- given the existing dataset -- it becomes overwhelmingly likely that
most of the lenses lie in a fattened LMC. This method, though, does
suffer from a possible bias if the Milky Way halo is under-endowed
with binaries.

In the longer term, a definitive test is to measure simultaneously the
photometric and astrometric microlensing signals of a few events with
the Space Interferometry Mission (SIM), which is currently scheduled
for launch in mid 2005. This suggestion has been advanced by Boden,
Shao \& van Buren (1998) and Gould \& Salim (1999). It enables the
unambiguous identification of the lens location at the cost of about 20
hours exposure time per event with SIM. Since this method is able to
discern the location of the lenses on an event-by-event basis, rather
than by ensemble likelihood statistics, SIM and SuperMACHO should
provide useful and complementary datasets. One way or another, the
location of the lenses will be known within five years or so.

\acknowledgments
We thank Ken Freeman, Andy Gould, Geza Gyuk, Paul Schechter and Will
Sutherland for a number of helpful conversations and suggestions.
Martin Weinberg and Pierre Salati kindly forwarded material in advance
of publication. NWE is supported by the Royal Society, while EK
acknowledges financial support from PPARC (grant number
GS/1997/00311).

\begin{table*}
\begin{center}
\begin{tabular}{ccccc}
Component & Model & Scale length & Scale height & Mass \\
\null & \null & [in kpc] & [in kpc] & [in $10^9 \msun$] \\
\hline
LMC disk & Double exponential & 1.6 & 0.3 & 4.0 \\
LMC shroud & Double exponential & 1.6 & 3.0 & 2.0  \\
LMC veil & Double exponential & 1.6 & 6.0 & 0.2 \\
LMC bar & Exponential spheroid & 1.6 & 0.3 & 0.4 \\
\end{tabular}
\end{center}
\caption{Description of the models used to represent the LMC in the
Monte Carlo simulation. The position angle of the bar is offset from
the position angle of the LMC disk by $50^\circ$. The overall mass in
the shroud and veil can be adjusted by $\gshroud$. In the Monte Carlo
simulations, $\gshroud$ is chosen so that the two competing models
have similar total numbers of events. Just the timescale and geometry
information are used to distinguish between them.}
\label{table:wyn}
\end{table*}
\begin{table*}
\begin{center}
\begin{tabular}{cccccccccccc}
 & \multicolumn{11}{c}{$\gshroud$} \\
\cline{2-12}
 & \multicolumn{3}{c}{1} & & \multicolumn{3}{c}{1 (no ang. dist.)} 
                                & & \multicolumn{3}{c}{0.5} \\
 & \multicolumn{3}{c}{$\fhalo$} & & \multicolumn{3}{c}{$\fhalo$} 
                               & & \multicolumn{3}{c}{$\fhalo$} \\
\cline{2-4} \cline{6-8} \cline{10-12}
$\mdark$ & 0.1 & 0.3 & 0.5 & & 0.1  & 0.3  & 0.5 & & 0.1 & 0.3 & 0.5 \\
\hline
$0.1~\msun$    & 9.5  & 2   & 1 & & 14 & 2   & 1 & & 6.5  & 1.5 & 1 \\
$0.5~\msun$    & 20.5 & 3.5 & 2 & & 35 & 4.5 & 2 & & 14.5 & 3   & 2 \\
\end{tabular}
\end{center}
\caption{Experiment lifetime $T$ (in years) required before $D_{95}$
exceeds 0.95 for various halo fractions, $\fhalo$, LMC
shroud mass factors, $\gshroud$, and halo/shroud lens masses, $m_{\rm
dark}$. For columns~5--7, headed ``(no ang. dist.)'', the
lifetimes are based on likelihood comparisons which ignore the angular
distribution of events.}
\label{table:eamonn}
\end{table*}
\begin{figure}
{
              \epsfxsize \hsize
               \leavevmode\epsffile{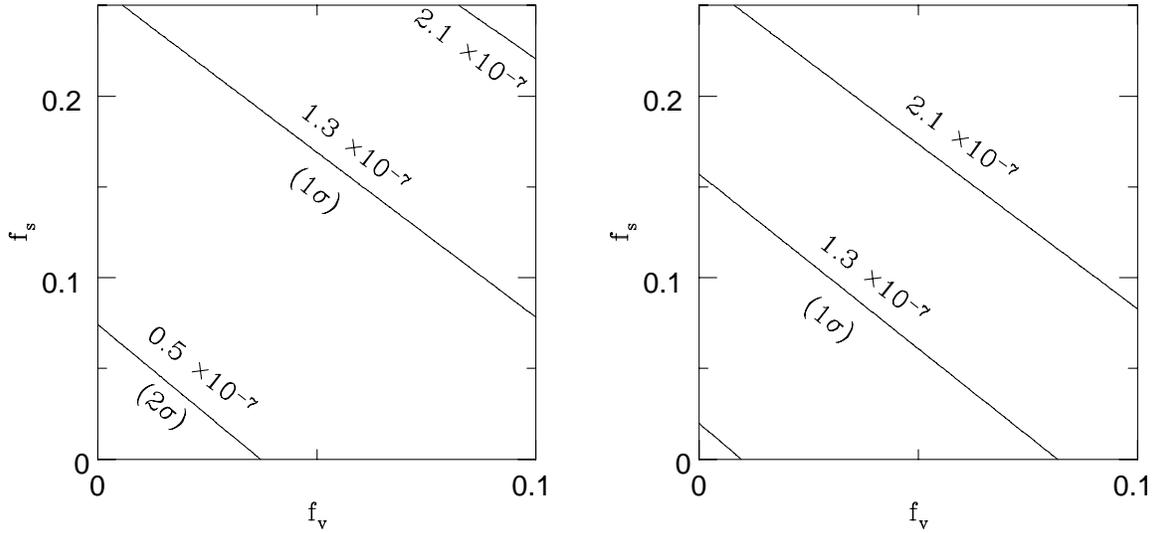}
}
\caption{Contours of the self-lensing optical depth of the LMC are
shown. Both panels show the effect of variation of the mass fractions
in the shroud $\fshroud$ and veil $\fv$, assuming $\hs/ \hd = 10$ and
$\hv/ \hd = 20$. The left panel assumes an inclination angle of
$27^\circ$, the right panel an inclination of $45^\circ$. The
observational estimate of the optical depth of $2.1 \times 10^{-7}$ is
shown, together with the $1\sigma$ and $2\sigma$ contours.}
\label{fig:contours}
\end{figure}
\begin{figure}
{
              \epsfxsize \hsize
               \leavevmode\epsffile{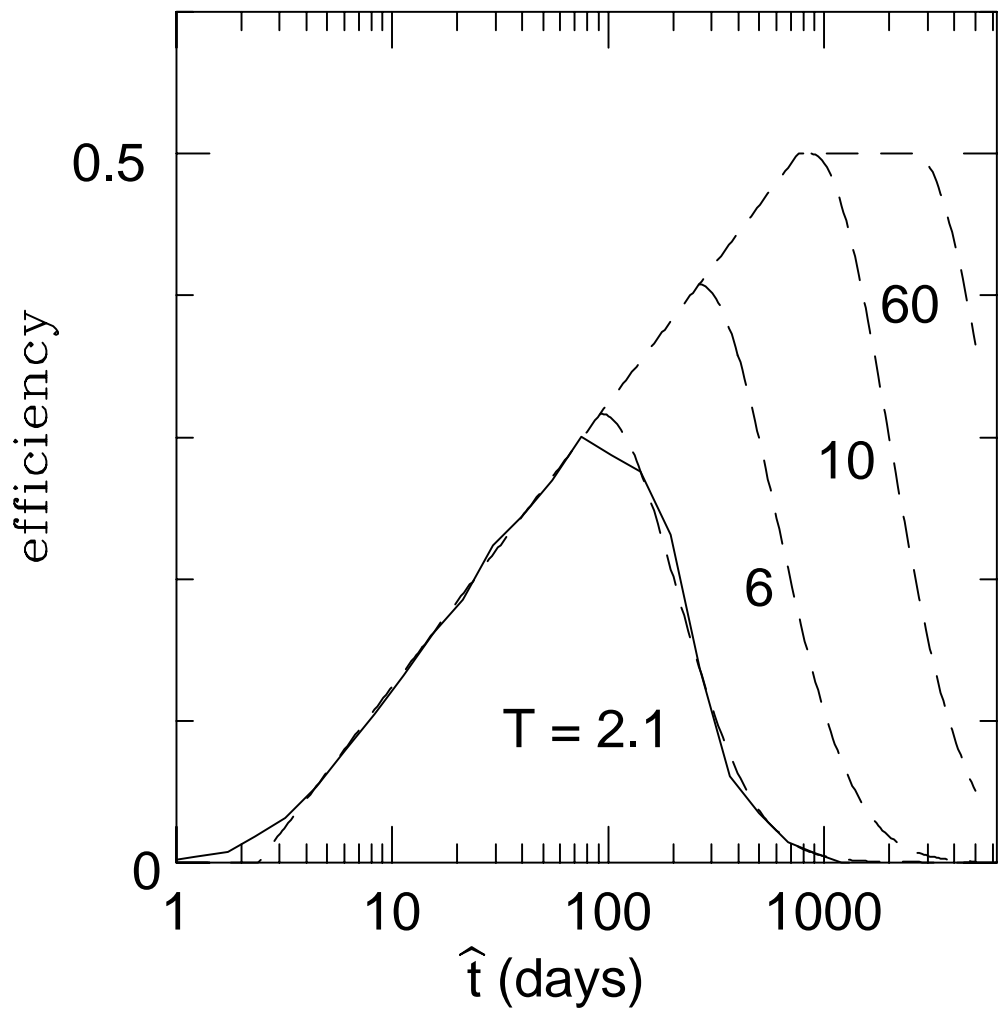}
}
\caption{The assumed evolution of detection efficiency towards the LMC
with experiment lifetime $T$ for $T = 2.1$, 6, 20 and 60 years (dashed
lines). The efficiency is assumed to increase with $T$ for longer
Einstein diameter crossing times $\that$ up to a maximum efficiency
level of 0.5. The actual detection efficiency of Alcock et al. (1997)
after 2.1 years observation is shown by the solid line.}
\label{fig:eff}
\end{figure}
\begin{figure}
\hspace*{2.5cm}
\rotate[r]{
              \epsfxsize 0.5\hsize
               \leavevmode\epsffile{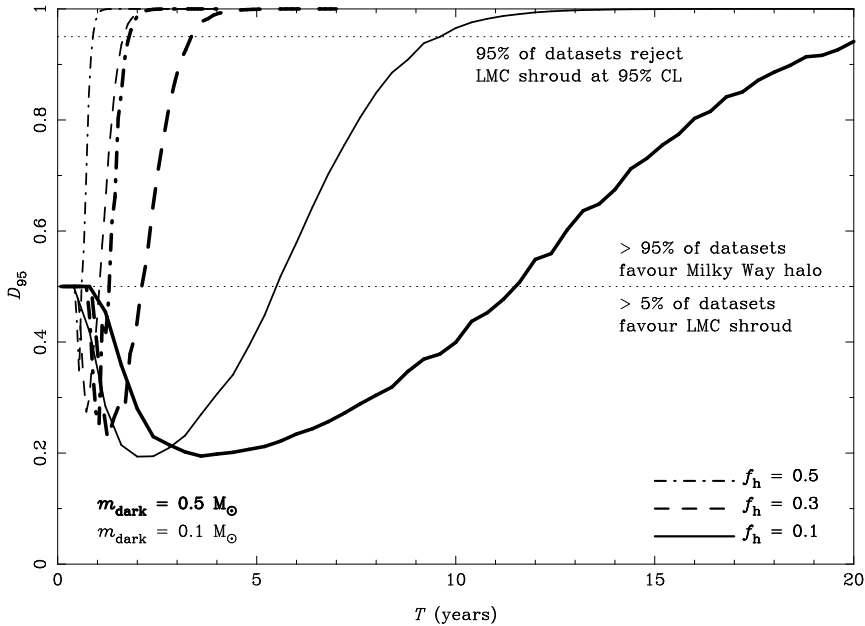}
}
\caption{The discriminatory power of our idealized LMC microlensing
experiment as a function of its lifetime $T$. The input (true) lens
model comprises an LMC disk and bar, together with a Milky Way halo
comprising MACHOs of mass $\mdark = 0.1~\msun$ (thin lines) and
$0.5~\msun$ (thick lines), and fractional contribution $\fhalo = 0.1$
(solid lines), 0.3 (dashed lines) and 0.5 (dot-dashed lines). The
comparison model consists of the same LMC disk and bar, but in place
of a Milky Way halo is a diffuse LMC shroud comprising lenses of the
same mass $\mdark$ but with a mass factor $\gshroud= 1$. The
discrimination measure $D$ represents the confidence with which the
data favours the input model (halo) over the comparison model (shroud)
after time $T$, and $D_{95}$ is its $95\%$ lower limit value derived
from a large ensemble of datasets. Lines dipping below the dotted line
at $D_{95} = 0.5$ indicate configurations in which more than $5\%$ of
simulated datasets misleadingly implicate the shroud model over the
halo model. Lines rising above the $D_{95} = 0.95$ dotted line
indicate that in $95\%$ of simulations sufficient data has been
accumulated to reject the shroud model with greater than $95\%$
confidence.}
\label{fig:res}
\end{figure}
\begin{figure}
\hspace*{2.5cm}
\rotate[r]{
              \epsfxsize 0.5\hsize
               \leavevmode\epsffile{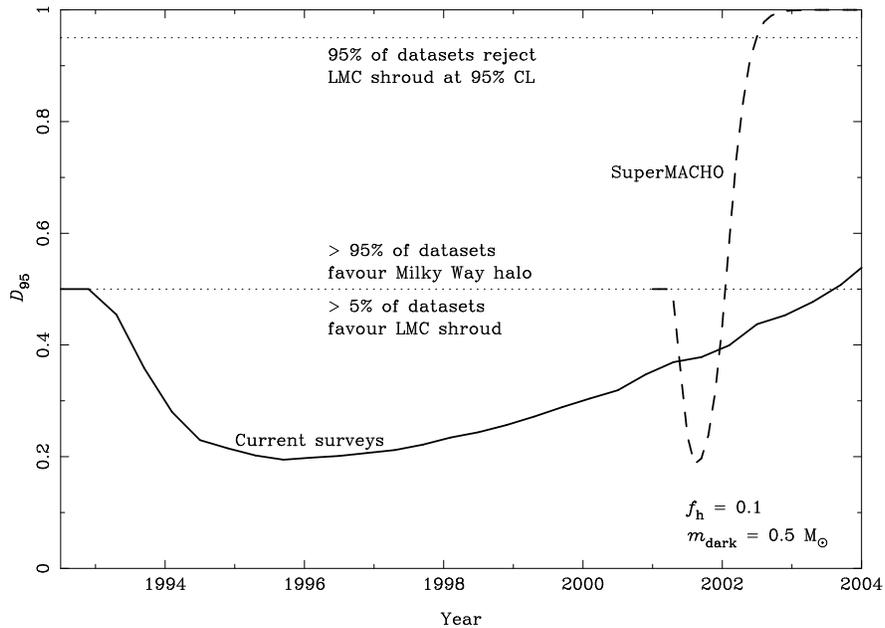}
}
\caption{The $95\%$ lower limit on the discriminatory power, $D_{95}$,
of current surveys (solid line) and the proposed next-generation
survey, ``SuperMACHO'', of Stubbs (1998) as a function of observation
epoch. We assume a starting date of 1992.5 for the current surveys and
2001 for SuperMACHO. The limits shown are for the halo model with
$\fhalo = 0.1$ and $\mdark = 0.5~\msun$.}
\label{fig:sm}
\end{figure}

\end{document}